\documentclass[12pt]{article}
\usepackage{amsfonts} 
\newcommand{\be}{\begin{equation}}
\newcommand{\ba}{\begin{eqnarray}}
\newcommand{\ee}{\end{equation}}
\newcommand{\ea}{\end{eqnarray}}
\newcommand{\cosech} { {\rm cosech}}

\begin{document}

\title{Rationally extended shape invariant potentials in arbitrary D-dimensions associated with exceptional $X_m$ polynomials}

\author{Rajesh Kumar Yadav$^{a}$\footnote{e-mail address: rajeshastrophysics@gmail.com (R.K.Y)}, Nisha Kumari$^{b}$\footnote{e-mail address: nishaism0086@gmail.com (N.K)}, Avinash Khare$^{c}$\footnote {e-mail address: khare@iiserpune.ac.in (A.K)} and\\
 Bhabani Prasad Mandal$^{b}$\footnote{e-mail address: bhabani.mandal@gmail.com (B.P.M).}}
 \maketitle
{$~^a$ Department of Physics, S. P. College, Dumka (SKMU Dumka)-814101, INDIA.\\
$~^b$ Department of Physics, Banaras Hindu University, Varanasi-221005, INDIA.\\ 
$~^c$Raja Ramanna Fellow, Indian Institute of Science Education and Research (IISER), Pune-411021, INDIA.}

\begin{abstract}
Rationally extended shape invariant potentials in arbitrary D-dimensions are obtained by using point 
canonical transformation (PCT) method. The bound-state solutions of these exactly solvable potentials 
can be written in terms of $X_m$ Laguerre or $X_m$ Jacobi exceptional orthogonal polynomials. These potentials 
are isospectral to their usual counterparts and possess translationally shape 
invariance property.

\end{abstract}

\section{Introduction}

 In recent years the discovery of exceptional orthogonal polynomials (EOPs) (also known as $X_1$ Laguerre and $X_1$ 
Jacobi polynomials) \cite{eop2,eop3} has increased the list of exactly solvable potentials. The EOPs are the solutions of second-
order Sturm-Liouvilli eigenvalue problem with rational coefficients. Unlike the usual orthogonal polynomials,
the EOPs starts with degree $n\geq 1$ and still form a complete orthonormal set with respect to a positive 
definite innerproduct defined over a compact interval. After the discovery of these two polynomials Quesne et.al. reported 
three shape invariant potentials whose solutions are in terms of $X_1$ Laguerre polynomial (extended radial oscillator 
potentials) and $X_1$ Jacobi polynomials (extended trigonometric scarf and generalized P\"oschl Teller (GPT) potentials) \cite{que,bqr}. 
Subsequently Odake and Sasaki generalize these and obtain the solutions in terms of exceptional $X_m$ orthogonal polynomials \cite{os}.
The properties of these $X_m$ exceptional orthogonal polynomials have been studied in detail in Ref. \cite{hos,hs,gom,qu}. 
Subsequently, the extension of other exactly solvable shape invariant potentials have also been done \cite{op1,op2,op3,op4} by using different 
approaches such as supersymmetry Quantum mechanics (SUSYQM) \cite{ew,cks}, point canonical transformation (PCT) \cite{pct}, Darboux-Crum transformation
(DCT) \cite{dt} etc. The scattering amplitude of some of the newly found exactly solvable potentials in 
terms of $X_m$ EOPs are studied in Ref. \cite{rabxm,rab,rabekrt}. The bound state solutions of these extended (deformed) potentials are in terms of EOPs or 
some type of new polynomials ($y_n$) ( which can be expressed in terms of combination of usual Laguerre or Jacobi orthogonal 
polynomials).

The bound state spectrum of all these extended potentials are investigated in a fixed dimension ($D=1$ or $3$).  Recently
the extension of some exactly solvable potentials have been made in arbitrary $D$ dimensions whose solutions are in terms
of $X_1$ EOPs \cite{dx1}. The obvious question then if one can extend this
discussion and obtain potentials whose solutions are in terms of $X_m$ EOPs
in arbitrary dimensions. The purpose of this paper is to answer this question. 
In particular, in this paper we apply the PCT approach, which consists of a 
coordinate transformation and a functional transformation, that allows 
generation of normalized exact analytic bound state solutions 
of the Schr\"odinger equation, starting from an analytically solvable 
conventional potential. Here we consider 
two analytically solved conventional potentials (they are isotropic oscillator and GPT potential) \cite{que,ks} corresponding to which $D$- dimensional rationally extended potentials are obtained whose solutions are in terms of $X_m$ exceptional 
Laguerre or Jacobi polynomials.  
    
This paper is organized as follow: In Sec. $2$, details about the point 
canonical transformation 
(PCT) method for arbitrary D-dimensions is given. In Sec. $3$, We have written the differential equation corresponding
to $X_m$ EOPs and discussed some important properties for EOPs. Arbitrary 
$D$-dimensional rationally extended exactly solvable potentials whose 
solutions are in terms of $X_m$ Laguerre or $X_m$ Jacobi EOPs are obtained in 
Sec. $4$. The approximate solutions corresponding to $X_m$ Jacobi 
case are also discussed in this section. In Sec. $5$, new 
shape invariant potentials  for the rationally extended $X_m$ Laguerre and $X_m$ Jacobi  
polynomials are obtained in arbitrary $D$ dimensions. In particular,  
for $D=2$ and  $D=4$, the shape invariant partner potentials for the extended 
radial oscillator are obtained explicitly. Sec. $6$, is reserved for 
results and discussions.

\section{Point canonical transformation (PCT) method for arbitrary D-dimensions}

In this section we discuss a more traditional approach, the PCT approach \cite{pct} to get the extension 
of conventional potentials by considering the radial Schr\"odinger equation in arbitrary D-dimensional 
Euclidean space \cite{kfd,es} given by $(\hbar=2m=1 )$

\be\label{rse}
\frac{d^2\psi(r)}{dr^2} + \frac{(D-1)}{r}\frac{d\psi(r)}{dr}+\bigg( E_n-V(r)-\frac{\ell(\ell + D-2)}{r^2}\bigg)\psi(r)=0.
\ee
To solve this equation we apply PCT approach and assume the solution of the form
\be\label{rse2}
\psi(r) = f(r)F(g(r)),
\ee
where $f(r)$ and $g(r)$ are two undetermined functions and $ F(g(r))$ will 
be later identified as one of the 
orthogonal polynomials which satisfies a second-order differential equation
\be\label{deq}
F''(g(r))+Q(g(r))F'(g(r))+R(g(r))F(g(r))=0.
\ee
Here a prime denote derivative with respect to $g(r)$.

Using Eq. (\ref{rse2}) in Eq.(\ref{rse}) and comparing the results with Eq. (\ref{deq}), we get
\be\label{fr}
f(r) = N\times r^{-\frac{(D-1)}{2}}(g'(r))^{-\frac{1}{2}}\exp\bigg(\frac{1}{2}\int{Q(g)dg}\bigg)
\ee
and
\be\label{ev}
E_n-V(r)-\frac{\ell(\ell + D-2)}{r^2} = \frac{1}{2}\{g(r),r\}+g(r)'^2\bigg (R(g)-\frac{1}{2}Q'(g)-\frac{1}{4}Q^2(g)\bigg)+\frac{(D-1)(D-3)}{4r^2},
\ee
where $N$ is the integration constant and plays the role of the normalization constant of the 
wavefunctions and $\{g(r),r\}$ is the Schwartzian derivative symbol \cite{he}, $\{g,r\}$ defined as
\be\label{he}
\{g(r),r\} = \frac{g'''(r)}{g'(r)}-\frac{3}{2}\frac{g''^2(r)}{g'^2(r)}.
\ee
Here the prime denotes derivative with respect to r.\\
From (\ref{rse2}) and (\ref{fr}), the normalizable wavefunction is given by
\be\label{wf}
\psi(r)=\frac{\chi(r)}{r^{\frac{(D-1)}{2}}},
\ee
where
\be\label{nwf}
\chi(r) = N\times (g'(r))^{-\frac{1}{2}}\exp\bigg(\frac{1}{2}\int{Q(g)dg}\bigg) F(g(r)).
\ee  
The radial wavefunction $\psi(r)=\frac{\chi (r)}{r^{(\frac{D-1}{2})}}$ has 
to satisfy the boundary condition $\chi(r)=0$ to  be more precise, it must 
at least vanish as fast as $r^{(D-1)/2}$ as $r$ goes to zero in order to 
rule out singular solutions \cite{an}.
For Eq. (\ref{ev}) to be satisfied, one needs to find some function $g(r)$ ensuring the presence of a constant term on its 
right hand side to compensate $E_n$ on its left hand one, while giving rise to a potential $V(r)$ with well behaved wavefunctions.
  
%It is interesting here to note that when we solve D-dimensional Schr\"odinger equation, the extended new potentials
% contain an extra term $\frac{(D-1)(D-3)}{4r^2}$, which bahaves as constant background 
%attractive inverse square potential in any arbitrary dimensions except for $D=1$ and $D=3$.    
%For power law cases, this background potential coming from Schwartzian derivative to give the correct
%centrifugal barrier potential in arbitrary dimensions.

\section{Exceptional $X_m$ orthogonal polynomials}

For completeness, we now give the differential equations corresponding to 
the $X_m$ EOPs and summarize the important properties for these two 
polynomials.

\subsection{Exceptional $X_m$ Laguerre orthogonal polynomials}

For an integer $m \geq  0$, $n\geq m$ and $k > m$, the $X_m $ Laguerre orthogonal polynomial $\hat{L}^{(\alpha)}_{n,m}(g(r))$ 
satisfy the differential equation \cite{xm1}
\ba\label{xmdeq}
\hat{L}^{''(\alpha)}_{n,m}(g(r))&+&\frac{1}{g}\bigg((\alpha+1-g)-2g\frac{L^{(\alpha)}_{m-1}(-g(r))}{L^{(\alpha-1)}_{m}(-g(r))}\bigg)\hat{L}^{'(\alpha)}_{n,m}(g(r))\nonumber\\
&+&\frac{1}{g}\bigg(n-2\alpha \frac{L^{(\alpha)}_{m-1}(-g(r))}{L^{(\alpha-1)}_{m}(-g(r))}\bigg)\hat{L}^{(\alpha)}_{n,m}(g(r))=0
\ea
The $\mathcal{L}^2$ norms of the $X_m$ Laguerre polynomials are given by
\be\label{norm_l}
\int^\infty_0\big(\hat{L}^{(\alpha)}_{n,m}(g)\big)^2 W^{\alpha}_m(g)dg=\frac{(\alpha+n)\Gamma(\alpha+n-m)}{(n-m)!},
\ee 
where
\be
W^{\alpha}_m(g)=\frac{g^{\alpha}e^{-g}}{(L^{(\alpha-1)}_{m}(-g))^2}
\ee
is the weight factor for the $X_m$ Laguerre polynomials.\\
In terms of classical Laguerre polynomials the $X_m$ Laguerre polynomials can be written as
\be
\hat{L}^{(\alpha)}_{n,m}(g)=L^{(\alpha)}_m(-g)L^{(\alpha-1)}_{n-m}(g)+L^{(\alpha-1)}_m(-g)L^{(\alpha)}_{n-m-1}(g); \quad n\geq m.
\ee
For $m=0$, the above definitions reduces to their classical counterparts i.e.
\be
\hat{L}^{(\alpha)}_{0,n}(g)=L^{(\alpha)}_{n}(g)
\ee 
\be
W^{\alpha}_0(g)={g^{\alpha}}{e^{-g}},
\ee
and for $m=1$ this satisfy Eq. (80) of Ref. \cite{eop2}. The other properties related to the $X_m$ Laguerre polynomials are discussed in detain in 
Ref. \cite{xm1}.

\subsection{Exceptional $X_m$ Jacobi orthogonal polynomials}

 For an integer $m\geq 1$ and $\alpha, \beta>-1$ the exceptional $X_m$ Jacobi orthogonal polynomials $\hat{P}^{(\alpha,\beta)}_{n,m}(g(r))$
\cite{xm2,bm} satisfies the differential equation
\ba\label{djacobi}
 \hat{P}^{''(\alpha,\beta)}_{n,m}(g(r))&+&\bigg((\alpha-\beta-m+1)\frac{P^{(-\alpha,\beta)}_{m-1}(g(r))}{P^{(-\alpha-1,\beta-1)}_{m}(g(r))}-\bigg(\frac{\alpha+1}{1-g(r)}\bigg)+\bigg(\frac{\beta+1}{1+g(r)}\bigg)\bigg)\hat{P}^{'(\alpha,\beta)}_{n,m}(g(r))\nonumber \\
 &+& \frac{1}{(1-g^2(r))}\bigg(\beta(\alpha-\beta-m+1)(1-g(r))\frac{P^{(-\alpha,\beta)}_{m-1}(g(r))}{P^{(-\alpha-1,\beta-1)}_{m}(g(r))}\nonumber \\
&+&m(\alpha-\beta-m+1)+(n-m)(\alpha+\beta+n-m+1)\bigg)\hat{P}^{(\alpha,\beta)}_{n,m}(g(r))=0 
\ea 
The $\mathcal{L}^2$ norms of the $X_m$ Jacobi polynomials are given by
\be
\int^{1}_{-1}[\hat{P}^{(\alpha,\beta)}_{n,m}(g(r))]^2 \hat{W}^{\alpha,\beta}_{m}dg=\frac{2^{\alpha+\beta+1}(1+\alpha+n-2m)(\beta+n)\Gamma(\alpha+2+n-m)\Gamma(\beta+n-m)}{(n-m)!(\alpha+1+n-m)^2(\alpha+\beta+2n-2m+1)\Gamma(\alpha+\beta+n-m+1)}
\ee
Where
\be
\hat{W}^{\alpha,\beta}_{m}=\frac{(1-g)^\alpha (1+g)^\beta}{[P^{(-\alpha-1,\beta-1)}_{m}(g(r))]^2}
\ee
is the weight factor for the $X_m$ Jacobi polynomials. The above $\mathcal{L}^2$ norms of the $X_m$ Jacobi polynomials holds, when 
the denominator of the above weight factor is non-zero for $-1\leq g\leq 1$. To ensure this, the following two conditions must be 
satisfied simultaneously:
\ba
&(i)& \beta\neq0, \quad \alpha, \alpha-\beta-m+1\not\in\{0,1,........,m-1\} \nonumber \\
&(ii)& \alpha>m-2, \mbox{sgn}(\alpha-m+1)=\mbox{sgn}(\beta),
\ea    
where $sgn(g)$ is the signum function.
In terms of classical Jacobi polynomials $P^{(\alpha,\beta)}_n(g)$, the $X_m$ Jacobi polynomials can be written as
\ba
\hat{P}^{(\alpha,\beta)}_{n,m}(g)=(-1)^m\bigg[\frac{1+\alpha+\beta+j}{2(1+\alpha+j)}(g-1)P^{(-\alpha-1,\beta-1)}_{m}(g)P^{(\alpha+2,\beta)}_{j-1}(g)\nonumber \\+\frac{1+\alpha-m}{\alpha+1+j}P^{(-2-\alpha,\beta)}_{m}(g)P^{(\alpha+1,\beta-1)}_{j}(g)\bigg];\quad j=n-m\geq 0.
\ea 
For $m=0$, the above definitions reduces to their classical counterparts i.e.
\be
\hat{P}^{(\alpha\beta)}_{0,n}(g)=P^{(\alpha,\beta)}_{n}(g)
\ee 
\be
W^{\alpha,\beta}_0(g)={(1-g)^{\alpha}}{(1+g)^\beta},
\ee
and for $m=1$ this satisfy Eq. (56) of Ref. \cite{eop2}. The other properties related to the $X_m$ Jacobi polynomials are discussed in detain in 
Ref. \cite{xm2}

\section{Extended potentials in $D$-dimensions}

\subsection{Potentials associated with $X_m$ exceptional Laguerre polynomial}
In this section we consider the extension of the usual radial oscillator potential. For this potential let us define 
the function $F(g)$ as an $X_m$ ($m\geq 1$) exceptional Laguerre polynomial ${\hat{L}^{(\alpha)}_{n,m}(g)}$, where $n=0,1,2,3,.....$, and
$\alpha > 0$, the associated second order differential Eq. (\ref{deq}) is equivalent to $X_m$ Laguerre differential 
equation (\ref{xmdeq}) where the functions $Q(g)$ and $R(g)$ are
\ba\label{qg}
Q(g)&=&\frac{1}{g}\bigg [(\alpha+1-g)-2g\frac{L^{(\alpha)}_{m-1}(-g)}{L^{(\alpha-1)}_{m}(-g)} \bigg] \nonumber \\
R(g)&=& \frac{1}{g} \bigg[n-2\alpha\frac{L^{(\alpha)}_{m-1}(-g)}{L^{(\alpha-1)}_{m}(-g)}\bigg]\,.
\ea
Using  $Q(g)$ and $ R(g)$ in Eq. (\ref{ev}), we get
\ba\label{eva}
E_n-V_m(r)&=&\frac{1}{2}\{ g,r\} + (g')^2\bigg( -\frac{1}{4} + \frac{n}{g}+\frac{(\alpha + 1)}{2g}-\frac{(\alpha+1)(\alpha-1)}{4g^2}+\frac{L^{(\alpha+1)}_{m-2}(-g)}{L^{(\alpha-1)}_{m}(-g)}\nonumber \\
&-&\frac{(\alpha+g-1)}{g}\frac{L^{(\alpha)}_{m-1}(-g)}{L^{(\alpha-1)}_{m}(-g)}-2\bigg(\frac{L^{(\alpha)}_{m-1}(-g)}{L^{(\alpha-1)}_{m}(-g)}\bigg)^2\bigg)+\frac{(D-1)(D-3)}{4r^2}.
\ea
To get $E_n$ as explained in the above section, here we assume $\frac{(g'(r))^2}{g(r)}=C_1$ (a constant not equal to zero), and for the 
radial oscillator potential this constant $C_1$ can be obtained by setting
\be\label{g}
g(r)=\frac{1}{4}C_1r^2.
\ee
Putting $g(r)$ in the above Eq. (\ref{eva}) and define the quantum number $n\rightarrow n+m$, we get
\be\label{ee}
E_n=nC_1;\qquad n=0,1,2,.....,
\ee
\ba\label{pot}
V_m(r)=\frac{1}{16}C_1^2r^2 + \frac{(\alpha+\frac{1}{2})(\alpha - \frac{1}{2})}{r^2}-\frac{C^2_1r^2}{4}\frac{L^{(\alpha+1)}_{m-2}(-g)}{L^{(\alpha-1)}_{m}(-g)}+C(\alpha+\frac{Cr^2}{4}-1)\nonumber\\
\times  \frac{L^{(\alpha)}_{m-1}(-g)}{L^{(\alpha-1)}_{m}(-g)}+\frac{c^2r^2}{2}\bigg(\frac{L^{(\alpha)}_{m-1}(-g)}{L^{(\alpha-1)}_{m}(-g)}\bigg)^2-\frac{C}{2}(2m+\alpha+1)-\frac{(D-1)(D-3)}{4r^2}.
\ea
The wavefunction can be obtained by putting $Q(g)$ and $g(r)$ in Eq.(\ref{nwf}) and is given by
\be\label{wfn}
\chi_{n,m}(r) = N_{n,m}\times \frac{r^{(\alpha+\frac{1}{2})}\exp\big(-\frac{C_1r^2}{8}\big)}{L^{(\alpha-1)}_{m}(-\frac{1}{4}Cr^2)}\hat{L}^{(\alpha)}_{n+m,m}\bigg(\frac{C_1r^2}{4}\bigg), 
\ee
where $N_{n,m}$ is the normalization constant given by
\be
N_{n,m}=\bigg(\frac{n!}{{(\alpha+n+m)}\Gamma(\alpha+n)}\bigg)^{\frac{1}{2}}.
\ee
To get the correct centrifugal barrier term in D-dimensional Euclidean space, we have to identify the coefficient of $\frac{1}{r^2}$ in Eq. (\ref{pot})
to be equal to $\ell(\ell+D-2)$, which fixes the value of $\alpha$ as 
\be
\alpha=\ell+\frac{D-2}{2}
\ee
and identifying the constant $C_1=2\omega$ , the energy eigenvalues (\ref{ee}), extended potential (\ref{pot}) and the corresponding 
wavefunction (\ref{wfn})  in any arbitrary $D$-dimensions are 
\be\label{ee2}
E_n=2n\omega
\ee
\ba\label{pot2}
V_{m}(r)= V^{D}_{\mbox{rad}}(r)&-&\omega^2r^2\frac{L^{(l+\frac{D}{2})}_{m-2}(-\frac{\omega r^2}{2})}{L_{m}^{(l+\frac{D-4}{2})}(-\frac{\omega r^2}{2})}+\omega(\omega r^2+2l+D-4)\frac{L^{(l+\frac{D-2}{2})}_{m-1}(-\frac{\omega r^2}{2})}{L_{m}^{(l+\frac{D-4}{2})}(-\frac{\omega r^2}{2})}\nonumber\\
&+&2\omega^2 r^2\bigg(\frac{L^{(l+\frac{D-2}{2})}_{m-1}(-\frac{\omega r^2}{2})}{L_{m}^{(l+\frac{D-4}{2})}(-\frac{\omega r^2}{2})}\bigg)^2 - 2m\omega)
\ea 
and
\be\label{wfn2}
\chi_{n,m}(r) = N_{n,m}\times \frac{r^{\ell+\frac{D-1}{2}}\exp\big(-\frac{\omega r^2}{4}\big)}{L^{(l+\frac{D-4}{2})}_m(-\frac{\omega r^2}{2})}\hat{L}_{n+m,m}^{(\ell+\frac{D-2}{2})}(\frac{\omega r^2}{2})
\ee 
respectively. Where $V_{\mbox{rad}}^{D}(r)=\frac{1}{4}\omega^2r^2 + \frac{\ell(\ell+D-2)}{r^2}-\omega (\ell+\frac{D}{2})$
is conventional radial oscillator potential in
arbitrary $D$-dimensional space \cite{sas}. Note that the full 
eigenfunction $\psi$ as given by Eq. (\ref{wf}) with $\chi$ as given by
Eq. (\ref{wfn2}).\\

For a check on our calculations, we now discuss few special cases of the results obtained in equations (\ref{pot2}) and (\ref{wfn2}).\\
 
{\bf Case (a): $m=0$} 

For $m=0$, from Eq. (\ref{pot2}) and (\ref{wfn2}) we get the well known usual radial oscillator potential in $D$-dimensions \cite{sas}, 
\be
V_{0}(r)=V_{\mbox{rad}}^{D}(r)=\frac{1}{4}\omega^2r^2 + \frac{\ell(\ell+D-2)}{r^2}-\omega (\ell+\frac{D}{2})
\ee
and the corresponding wavefunctions which can be written in terms of usual Laguerre polynomials
\be
\chi_{n,0}(r) = N_{n,0}\times r^{\ell+\frac{D-1}{2}}\exp\big(-\frac{\omega r^2}{4}\big) L_{n}^{(\ell+\frac{D-2}{2})}(\frac{\omega r^2}{2}).
\ee 
For $D=3$ above expressions reduces to the well known 3-D harmonic oscillator potential.\\

{\bf Case (b): $m=1$}

 For $m=1$, the obtained potential
\be
V_{1}(r)=\frac{1}{4}\omega^2r^2 + \frac{\ell(\ell+D-2)}{r^2}-\omega (\ell+\frac{D}{2})+\frac{4\omega}{(\omega r^2+2\ell+D-2)}-\frac{8\omega(2\ell+D-2)}{(\omega r^2+2\ell+D-2)^2}
\ee
is the rationally extended $D$ dimensional oscillator potential studied earlier in Ref. \cite{dx1}\\
The corresponding wavefunctions in terms of exceptional $X_1$ Laguerre orthogonal polynomial can be written as
 \be
\chi_{n,1}(r) = N_{n,1}\times \frac{r^{\ell+\frac{D-1}{2}}\exp\big(-\frac{\omega r^2}{4}\big)}{(\omega r^2+2\ell+D-2)} \hat{L}_{n+1,1}^{(\ell+\frac{D-2}{2})}(\frac{\omega r^2}{2}).
\ee  
 For $D=3$, the above expressions matches exactly with the expressions given in Ref. \cite{que}.\\

{\bf Case (c): $m=2$} 

For $m=2$, In this case the extended potential and the corresponding wavefunctions in terms of $X_2$ Laguerre
orthogonal polynomials are given by
\ba
V_{2}(r)=\frac{1}{4}\omega^2r^2 &+& \frac{\ell(\ell+D-2)}{r^2}-\omega (\ell+\frac{D}{2})\nonumber \\
&+&\frac{8\omega[\omega r^2-(2\ell+D)]}{[\omega^2r^4+2\omega r^2(2\ell+D)+(2\ell+D-2)(2\ell+D)]}\nonumber \\
&+&\frac{64\omega^2r^2(2\ell+D)}{[\omega^2r^4+2\omega r^2(2\ell+D)+(2\ell+D-2)(2\ell+D)]^2},
\ea
and
\be
\chi_{n,2}(r) = N_{n,2}\times \frac{r^{\ell+\frac{D-1}{2}}\exp\big(-\frac{\omega r^2}{4}\big)}{(\omega^2r^4+2\omega r^2(2\ell+D)+(2\ell+D-2)(2\ell+D))} \hat{L}_{n+2,2}^{(\ell+\frac{D-2}{2})}(\frac{\omega r^2}{2}).
\ee   
\subsection{Potentials associated with $X_m$ exceptional Jacobi polynomial}

Let us consider the case where the second order differential equation (\ref{deq}) coincides with that satisfied by $X_m$ Jacobi polynomial $\hat{P}_{n,m}^{(\alpha,\beta)}$, where $n=1,2,3,.....; m\geq 1; \alpha,\beta > -1$ and $\alpha\not=\beta$.  
Thus the function $F(g)$ in Eq. (\ref{deq}) is equivalent to $\hat{P}_{n,m}^{(\alpha,\beta)}(g)$ and the other two functions $Q(g)$ and $R(g)$
are given by Eq. (\ref{djacobi})
\ba\label{q}
Q(g)&=&(\alpha-\beta-m-1)\frac{P^{(-\alpha,\beta)}_{m-1}(g)}{P^{(-\alpha-1,\beta-1)}_{m}(g)}-\frac{\alpha+1}{1-g}+\frac{\beta+1}{1+g}\nonumber\\
R(g)&=&\frac{\beta(\alpha-\beta-m+1)}{1+g}\frac{P^{(-\alpha,\beta)}_{m-1}(g)}{P^{(-\alpha-1,\beta-1)(g)}_{m}}\nonumber\\
&+&\frac{1}{1-g^2}\bigg((\alpha-\beta-m+1)+(n-m)(\alpha+\beta+n-m+1)\bigg).
\ea
Using the above equations in Eq. (\ref{ev}) and after doing some straightforward calculations for s-wave ($\ell=0$), we get
\ba\label{ev2}
E_n-V_{eff,m}(r)&=&\frac{1}{2}\{g(r),r\}+\frac{1-\alpha^2}{4}\frac{g'(r)^2}{(1-g(r))^2}+\frac{1-\beta^2}{4}\frac{g'(r)^2}{(1+g(r))^2}\nonumber\\
&+&\frac{2n^2+2n(\alpha+\beta-2m+1)+2m(\alpha-3\beta-m+1)+(\alpha+1)(\beta+1)}{2}\frac{g'(r)^2}{1-g(r)^2}\nonumber\\
&+&\frac{(\alpha-\beta-m+1)(\alpha+\beta+(\alpha-\beta+1)g(r))g'(r)^2}{1-g(r)^2}\frac{P^{(-\alpha,\beta)}_{m-1}(g)}{P^{(-\alpha-1,\beta-1)}_{m}(g)}\nonumber\\
&-&\frac{(\alpha-\beta-m+1)^2g'(r)^2}{2}\bigg(\frac{P^{(-\alpha,\beta)}_{m-1}(g)}{P^{(-\alpha-1,\beta-1)}_{m}(g)}\bigg)^2,
\ea
%Where
%\ba\label{cons}
%C&=&\frac{(\beta-\alpha)(\beta+\alpha)}{2\alpha\beta}, \qquad D=n^2+(\beta+\alpha-1)n + \frac{1}{4}[(\beta+\alpha)^2-2(\beta+\alpha)-4]+\frac{\beta^2+\alpha^2}{2\alpha\beta},\nonumber \\ 
%G&=&\frac{1}{2}(\beta-\alpha )(\beta+\alpha ),\qquad J = -\frac{1}{2}(\beta^2+\alpha^2-2)
%\ea
and the wavefunction (\ref{nwf}) becomes   
\be\label{wfn3}
\chi_{n,m}(r) = N_{n,m}\times g'(r)^{-\frac{1}{2}}\frac{(1+g)^{(\frac{\beta+1}{2})}(1-g)^{(\frac{\alpha+1}{2})}}{P^{(-\alpha-1,\beta-1)}_m(g)}\hat{P}_{n,m}^{(\alpha,\beta)}(g), 
\ee 
where the effective potential, $V_{eff,m}(r)$ is given by
\be\label{veff}
V_{eff,m}(r)=V_{m}(r)+\frac{(D-1)(D-3)}{4r^2},
\ee
and the normalization constant 
\be
N_{n,m}=\bigg(\frac{n!(\alpha +n+1)^2(\alpha +\beta +2n+1)\Gamma(\alpha+\beta+n+1)}{2^{\alpha+\beta+1}(1+\alpha +n-m)(\beta+n+m)\Gamma(\alpha+n+2)\Gamma(\beta+n)}\bigg)^{\frac{1}{2}}.
\ee 

It is interesting here to note that when this extended potential is purely non-power law, the potential given by Eq.(\ref{veff}) has an extra term 
$\frac{(D-1)(D-3)}{4r^2} $ which behaves as constant background attractive inverse square potential in any arbitrary dimensions except for 
$D=1$ or $3$. For power law cases (e.g. radial oscillator potential), this background potential give the correct barrier potential in arbitrary
dimensions (as shown in Eq. (\ref{pot2})). 

On using Eq. (\ref{he}) in Eq.(\ref{ev2}) and assume a term $\frac{g'^2(r)}{1-g^2(r)}=C_2$ (a real constant not equal to zero), we get a constant term on right hand side which gives the energy eigenvalue $E_n$. There are several possibilities of $g(r)$ which produces this constant $C_2$.\\ 
If we consider $g(r)=\cosh r$ ; $0\leq r\leq \infty $ and define the parameters $\alpha=B-A-\frac{1}{2}, \beta=-B-A-\frac{1}{2}$; $B>A+\frac{(D-1)}{2}>\frac{(D-1)}{2}$ and the quantum number $n\rightarrow n+m; m\geq 1$, 
 Eq. (\ref{ev2}) and Eq. (\ref{wfn3}) gives, 
 \be\label{ee3}
 E_n= -(A-n)^2,\qquad n=0,1,2,......,n_{max} \qquad A-1\leq n_{max}<A,
 \ee
 \ba\label{pot3}
 V_{eff,m}(r)&=&V_{m}(r)+\frac{(D-1)(D-3)}{4r^2}\nonumber \\
&=&V_{GPT}(r)+2m(2B-m+1)-(2B-m+1)[(2A+1-(2B+1)\cosh r)]\nonumber \\
&\times & \frac{P_{m-1}^{(-\alpha,\beta)}(\cosh r)}{P_{m}^{(-\alpha-1,\beta-1)}(\cosh r)}
 +\frac{(2B-m+1)^2\sinh^2 r}{2}\bigg(\frac{P_{m-1}^{(-\alpha,\beta)}(\cosh r)}{P_{m}^{(-\alpha-1,\beta-1)}(\cosh r)}\bigg)^2 
\ea
 and the wave function 
\be\label{wfn4}
 \chi_{n,m}(r) = N_{n,m}\times \frac{(\cosh r-1)^{(\frac{B-A}{2})}(\cosh r+1)^{-(\frac{B+A}{2})}}{P_{m}^{(-B+A-\frac{1}{2},-B-A-\frac{3}{2})}(\cosh r)}\hat{P}_{n+m,m}^{(B-A-\frac{1}{2},-B-A-\frac{1}{2})}(\cosh r). 
 \ee
 Where
\be\label{gpt} 
V_{GPT}(r) = (B^2+A(A+1))  \cosech^2 r-B(2A+1)  \cosech r  \coth r 
\ee
 is the conventional generalized P\"oschl Teller (GPT) potential. Note that
the full eigenfunction is given by Eq. (\ref{wf}) with $\chi$ as given by
Eq. (\ref{wfn4}).\\
Here we see that the energy eigenvalues of conventional potentials are same as the rationally extended D-dimensional potentials (i.e they are isospectral). \\

It is interesting to note that compared to $D=3$, the only change in the 
potential in D-dimensions is the extra centrifugal barrier term 
$\frac{(D-1)(D-3)}{4r^2}$, and $\chi(r)$ is 
unaltered while only $\psi(r)$ is slightly different due to $r^{(D-1)/2}$. 
It is also worth pointing out that even in three dimensions, the GPT
potential (\ref{gpt}) can be analytically solved only in the case of $S$-wave,
i.e. $l=0$. 

We now show that approximate solution of the GPT potential problem for 
arbitrary $l$ can, however, be obtained in $D$ dimensions.

\subsubsection{Approximate solutions for arbitrary $\ell$ }

In this section, we solve the $D$-dimensional Sch\"odringer equation (\ref{rse}) with arbitrary $\ell$ and obtain the 
effective potential (as given in Eq. (\ref{ev2})) with an extra $\ell$ dependent term i.e., 
\be\label{veff2}
V_{eff,m}(r)=V_{m}(r)+\bigg(\frac{(D-1)(D-3)}{4r^2}+\frac{\ell(\ell+D-2)}{r^2}\bigg).
\ee
So, in order to get the appropriate centrifugal barrier terms in the above 
effective potential, we have to apply some approximation. Following 
\cite{approx}, we consider the approximation 
\be
\frac{1}{r^2}\simeq \frac{1}{\sinh^2 r}\,.
\ee
Thus effectively, one has approximated a problem for the $l$'th partial wave 
to that of $l=0$ but with different set of parameters compared to the usual
$l=0$ case.
In that case, the effective potential (\ref{veff}) becomes
\be\label{veff3}
V_{eff,m}=V_{m}(r)+\bigg(\frac{(D-1)(D-3)}{4}+\ell(\ell+D-2)\bigg) \rm {cosech^2r}.
\ee 
Now we define the parameters $\alpha$ and $\beta$ in terms of modified parameters\footnote{When we solve the Schr\"odinger equation for usual GPT potential, $V_{GPT}(r)= (B^2+A(A+1)){\rm cosech^2 r}-B(2A+1)\coth r {\rm cosech r}$,
we define the parameters $\alpha$ and $\beta$ in terms of $A$ and $B$. But, in the above D-dimensional effective potential the parameters $\alpha$ and $\beta$
have been modified due to the presence of an extra D-dependent term.} $B'$ and $A'$ i.e., $\alpha=B'-A'-\frac{1}{2}, \beta=-B'-A'-\frac{1}{2}$;
$B'>A'+\frac{(D-1)}{2}>\frac{(D-1)}{2}$,\\
 where
\be\label{b'}
B'=\bigg[\frac{(\zeta+\frac{1}{4})+\big[(\zeta+\frac{1}{4}+B(2A+1))(\zeta+\frac{1}{4}-B(2A+1))\big]^{\frac{1}{2}}}{2}\bigg]^{\frac{1}{2}}
\ee
and
\be\label{a'}
A'=\frac{1}{2}\bigg[\frac{2B(A+\frac{1}{2})}{B'}-1\bigg],
\ee
while
\be\label{zeta}
\zeta=B^2+A(A+1)+\ell(\ell+D-2)+\frac{(D-1)(D-3)}{4}.
\ee
For $D=3$ and $\ell=0$; we get the usual parameters as defined in the above section i.e., $B'\rightarrow B$ and $A'\rightarrow A$.
On using these new parameters $\alpha$ and $\beta$, quantum number $n\rightarrow n+m; m\geq 1$, 
 Eq. (\ref{ev2}) and Eq. (\ref{wfn3}) gives, 
 \be\label{ee4}
 E_n= -(A'-n)^2,\qquad n=0,1,2,......,n_{max} \qquad A'-1\leq n_{max}<A',
 \ee
 \ba\label{pot4}
 V_{eff,m}(r)&=&V^{(A',B')}_{GPT}(r)+2m(2B'-m+1)-(2B'-m+1)[(2A'+1-(2B'+1)\cosh r)]\nonumber \\
&\times & \frac{P_{m-1}^{(-\alpha,\beta)}(\cosh r)}{P_{m}^{(-\alpha-1,\beta-1)}(\cosh r)}
 +\frac{(2B'-m+1)^2\sinh^2 r}{2}\bigg(\frac{P_{m-1}^{(-\alpha,\beta)}(\cosh r)}{P_{m}^{(-\alpha-1,\beta-1)}(\cosh r)}\bigg)^2 
\ea
 and the wave functions 
\be\label{wfn5}
 \chi_{n,m}(r) = N_{n,m}\times \frac{(\cosh r-1)^{(\frac{B'-A'}{2})}(\cosh r+1)^{-(\frac{B'+A'}{2})}}{P_{m}^{(-B'+A'-\frac{1}{2},-B'-A'-\frac{3}{2})}(\cosh r)}\hat{P}_{n+m,m}^{(B'-A'-\frac{1}{2},-B'-A'-\frac{1}{2})}(\cosh r). 
 \ee
 Where
\be 
V^{(A',B')}_{GPT}(r) = (B'^2+A'(A'+1))  \cosech^2 r-B'(2A'+1)  \cosech r  \coth r 
\ee
 is the conventional generalized P\"oschl Teller (GPT) potential in arbitrary $D$ and $\ell$.\\
 
 Similar to the extended oscillator case, we now consider few special cases 
of the results of extended GPT potential 
obtained in equations (\ref{pot4}) and (\ref{wfn5}).  

{\bf Case (a): $m=0$}

 For $m=0$, from Eq. (\ref{pot4}) and (\ref{wfn5}), the potential and the corresponding wavefunctions in terms 
of usual Jacobi polynomials are  
\be
V_{eff,0}(r)=V^{(A',B')}_{GPT}(r)
\ee
and 
\be
\chi_{n,0}(r) = N_{n,0}\times (\cosh r-1)^{(\frac{B'-A'}{2})}(\cosh r+1)^{-(\frac{B'+A'}{2})}{P}_{n}^{(B'-A'-\frac{1}{2},-B'-A'-\frac{1}{2})}(\cosh r). .
\ee 
For $D=3$ and $\ell=0$ the effective potential $V_{eff,0}=V_{GPT}(r)$.\\ \\ \\

{\bf Case (b): $m=1$}

For $m=1$, the obtained potential
\ba
V_{eff,1}(r)&=&V^{(A',B')}_{GPT}(r)+\frac{2(2A'+1)}{(2B'\cosh r-2A'-1)}-\frac{2[4B'^2-(2A'+1)^2]}{(2B'\cosh r-2A'-1)^2}
\ea
is the rationally extended $D$ dimensional GPT potential.\\ 
The corresponding wavefunctions in terms of exceptional $X_1$ Jacobi orthogonal polynomials can be written as
\be
\chi_{n,1}(r) = N_{n,1}\times \frac{(\cosh r-1)^{(\frac{B'-A'}{2})}(\cosh r+1)^{-(\frac{B'+A'}{2})}}{(2B'\cosh r-2A'-1)}\hat{P}_{n+1,1}^{(B'-A'-\frac{1}{2},-B'-A'-\frac{1}{2})}(\cosh r). 
\ee
 For $D=3$ and $\ell=0$, the above expressions matches exactly with the results obtained in \cite{bqr,rab}. 
 
 {\bf Case (c): $m=2$}

In this case the potential and its wavefunctions in terms of $X_2$ Jacobi polynomials are given by 
\ba
V_{eff,2}(r)&=&V^{(A',B')}_{GPT}(r)+4(2B'-1)\nonumber \\
&-&\frac{4[3(2B'-1)(2A'+1)\cosh r-2B'(2B'-1)-8A'(A'+1)]}{[(2B'-1)(2B'-2)\cosh^2 r-2(2B'-1)(2A'+1)\cosh r+4A'(A'+1)+2B'-1]}\nonumber \\
&+&\frac{8(2B'-1)^2\sinh^2 r [(2A'+1)-(2B'-2)\cosh r]^2}{[(2B'-1)(2B'-2)\cosh^2 r-2(2B'-1)(2A'+1)\cosh r+4A'(A'+1)+2B'-1]^2}\nonumber\\
&-&8
\ea
and\\
\ba
\chi_{n,2}(r)=&N_{n,2}&\frac{(\cosh r-1)^{(\frac{B'-A'}{2})}(\cosh r+1)^{-(\frac{B'+A'}{2})}}{[(2B'-1)(2B'-2)\cosh^2 r-2(2B'-1)(2A'+1)\cosh r+4A'(A'+1)+2B'-1]}\nonumber\\
&\times &\hat{P}_{n+2,2}^{(B'-A'-\frac{1}{2},-B'-A'-\frac{1}{2})}(\cosh r). 
\ea

 \section{New shape invariant potentials (SIPs) in higher dimensions}
 
 In supersymmetric quantum mechanics (SUSYQM) \cite{ew,cks} the superpotential $W(x)$ determines the two-partner potentials 
 \be
 V^{\pm}(x)=W^2(x)\pm W'(x)+E_{0}; \qquad \hbar=2m=1,
 \ee
 where $E_{0}$ is factorization energy\footnote{For radial oscillator potential the factorization energy $E_{0}=\omega(l+\frac{D}{2})$ and 
for GPT potential $E_{0}=-A^2$.}.\\
 For unbroken SUSY, these partner potentials satisfy a shape invariant property
\be\label{sip}
V^{(+)}(x; a_1) = V^{(-)}(x; a_2) + R(a_1), 
\ee
where $a_1$ is a set of parameters, $a_2$ is a function of $a_1$ (say $a_2 = f(a_1)$) and the remainder
$R(a_1)$ is independent of $x$.\\
 The eigenstates of these partner potentials are related by 
\be
E^{(+)}_{n}=E^{(-)}_{n+1} \qquad E^{(0)}_{0}=0; 
\quad \psi^{(+)}_{n}\propto A\psi^{(-)}_{n+1}
\qquad  \psi^{(-)}_{n+1}\propto A^{\dagger}\psi^{(+)}_{n},
\ee
Where $A$, $A^{\dagger }$ and superpotential $W(x)$ are defined as
\be
A=\frac{d}{dx}+W(x), \qquad   A^{\dagger}=-\frac{d}{dx}+W(x), 
\qquad W(x)=-\frac{d}{dx}[\ln\psi^{(-)}_{0}(x)].
\ee
The factorized Hamiltonians in terms of $A$ and $A^{\dagger}$ or in terms of partner potentials $V^{(\pm)}$ are given by
\be
H^{(-)}=A^{\dagger}A=-\frac{d^2}{dx^2}+V^{(-)}(x)-E, 
\qquad   H^{(+)}=AA^{\dagger}=-\frac{d^2}{dx^2}+V^{(+)}(x)-E.
\ee 

\subsection{Extended radial oscillator potentials}

We now show that the extended radial oscillator potentials that we have 
obtained in $D$ dimensions in Sec. 4 provide us with yet another example of
shape invariant potentials with translation.

For the radial oscillator case the ground state wave function $\chi^{(-)}_{0,m}(r)$ is given by Eq. (\ref{wfn2}) i.e.
\be
\chi^{(-)}_{0,m}(r)\propto \phi_{0}(r)\phi_{m}(r),
\ee
where
\be
\phi_{0}(r)\propto r^{l+\frac{D-1}{2}}\exp (-\frac{\omega r^2}{4}) \quad \mbox{and} \quad \phi_m(r)\propto \frac{L^{(l+\frac{D-2}{2})}_{m}(-\frac{\omega r^2}{2})}{L^{(l+\frac{D-4}{2})}_{m}(-\frac{\omega r^2}{2})}. 
\ee 
Here we see that the ground state wave function of the extended radial oscillator potential in higher dimensions whose 
solutions are in terms of EOPs differs from that of the usual potential by an extra term $\phi_{m}(r)$ and  
the corresponding superpotential $W(r)(=-\frac{d}{dr}[\ln\chi^{(-)}_{0,m}(r)])$ is given by 
 \be
 W(r)=W_{1}(r)+W_{2}(r),
 \ee
 where 
 \be
 W_{1}(r)=-\frac{\phi'_{0}(r)}{\phi_{0}(r)} \quad \mbox{and} \quad  W_{2}(r)=-\frac{\phi'_{m}(r)}{\phi_{m}(r)}.
 \ee
 Using $W(r)$, we get $V^{(-)}_{m}(r)( = W(r)^2-W'(r))$ same as in 
Eq. (\ref{pot2}) and $V^{(+)}_{m}(r)( = W(r)^2+W'(r))$
 is given by
\ba\label{rad2}
V^{(+)}_{m}(r)= V^{D,l+1}_{\mbox{rad}}(r)&-&\omega^2r^2\frac{L^{(l+\frac{D+2}{2})}_{m-2}(-\frac{\omega r^2}{2})}{L_{m}^{(l+\frac{D-2}{2})}(-\frac{\omega r^2}{2})}+\omega(\omega r^2+2l+D-2)\frac{L^{(l+\frac{D}{2})}_{m-1}(-\frac{\omega r^2}{2})}{L_{m}^{(l+\frac{D-2}{2})}(-\frac{\omega r^2}{2})}\nonumber\\
&+&2\omega^2 r^2\bigg(\frac{L^{(l+\frac{D}{2})}_{m-1}(-\frac{\omega r^2}{2})}{L_{m}^{(l+\frac{D-2}{2})}(-\frac{\omega r^2}{2})}\bigg)^2 - 2m\omega-(l+\frac{D-2}{2})
\ea
 From the above equations (\ref{pot2}) and (\ref{rad2}), the potential $V^{(+)}_{m}(r)$ can be obtained directly by replacing $l\longrightarrow l+1$ 
in $V^{(-)}_{m}(r)$ and satisfy Eq. (\ref{sip}). This means these two partner potentials are shape invariant potentials (with translation).\\
Thus we see that the same oscillator potential $V(r)=\frac{1}{4}\omega^2r^2$, where $r=\sqrt{x^2_1+x^2_2+......+x^2_D}$, gives different SIPs in different 
dimensions. For example:\\
For $D=2$
\ba
V^{(-)}_{m}(r)= \frac{1}{4}\omega^2 r^2+\frac{l^2}{r^2}&-&\omega^2r^2\frac{L^{(l+1)}_{m-2}(-\frac{\omega r^2}{2})}{L_{m}^{(l-1)}(-\frac{\omega r^2}{2})}+\omega(\omega r^2+2l-2)\frac{L^{(l)}_{m-1}(-\frac{\omega r^2}{2})}{L_{m}^{(l-1)}(-\frac{\omega r^2}{2})}\nonumber\\
&+&2\omega^2 r^2\bigg(\frac{L^{(l)}_{m-1}(-\frac{\omega r^2}{2})}{L_{m}^{(l-1)}(-\frac{\omega r^2}{2})}\bigg)^2 - 2m\omega-\omega(l+1)
\ea
and
\ba 
V^{(+)}_{m}(r)= \frac{1}{4}\omega^2 r^2+\frac{(l+1)}{r^2}&-&\omega^2 r^2\frac{L^{(l+2)}_{m-2}(-\frac{\omega r^2}{2})}{L_{m}^{(l)}(-\frac{\omega r^2}{2})}+\omega(\omega r^2+2l)\frac{L^{(l+1)}_{m-1}(-\frac{\omega r^2}{2})}{L_{m}^{(l)}(-\frac{\omega r^2}{2})}\nonumber\\
&+&2\omega^2 r^2\bigg(\frac{L^{(l+1)}_{m-1}(-\frac{\omega r^2}{2})}{L_{m}^{(l)}(-\frac{\omega r^2}{2})}\bigg)^2 - 2m\omega-\omega l
\ea
For $D=4$ 
\ba
V^{(-)}_{m}(r)= \frac{1}{4}\omega^2 r^2+\frac{l(l+2)}{r^2}&-&\omega^2 r^2\frac{L^{(l+2)}_{m-2}(-\frac{\omega r^2}{2})}{L_{m}^{(l)}(-\frac{\omega r^2}{2})}+\omega(\omega r^2+2l)\frac{L^{(l+1)}_{m-1}(-\frac{\omega r^2}{2})}{L_{m}^{(l)}(-\frac{\omega r^2}{2})}\nonumber\\
&+&2\omega^2 r^2\bigg(\frac{L^{(l+2)}_{m-1}(-\frac{\omega r^2}{2})}{L_{m}^{(l)}(-\frac{\omega r^2}{2})}\bigg)^2 - 2m\omega-\omega(l+2)
\ea
and
\ba
V^{(+)}_{m}(r)= \frac{1}{4}\omega^2 r^2+\frac{(l+1)(l+3)}{r^2}&-&\omega^2r^2\frac{L^{(l+3)}_{m-2}(-\frac{\omega r^2}{2})}{L_{m}^{(l+1)}(-\frac{\omega r^2}{2})}+\omega(\omega r^2+2l+2)\frac{L^{(l+2)}_{m-1}(-\frac{\omega r^2}{2})}{L_{m}^{(l+1)}(-\frac{\omega r^2}{2})}\nonumber\\
&+&2\omega^2 r^2\bigg(\frac{L^{(l+2)}_{m-1}(-\frac{\omega r^2}{2})}{L_{m}^{(l+1)}(-\frac{\omega r^2}{2})}\bigg)^2 - 2m\omega-(l+1).
\ea

\subsection{Extended P\"oschl-Teller potentials}

Let us first discuss the extended GPT (with $l=0$) as discussed in Sec. 4.2. 
In this case, the ground state wave 
function $\chi^{(-)}_{0,m}(r)$ is given by Eq. (\ref{wfn4}) i.e.
\be
\chi^{(-)}_{0,m}(r)\propto \phi_{0}(r)\phi_{m}(r)
\ee
Where
\be
\phi_{0}(r)\propto (\cosh r-1)^{(\frac{B-A}{2})}(\cosh r+1)^{-(\frac{B+A}{2})} \quad \mbox{and}\quad \phi_{m}(r)\propto \frac{P^{(-B+A-\frac{3}{2},-B-A-\frac{1}{2})}_{m}(\cosh r)}{P^{(-B+A-\frac{1}{2},-B-A-\frac{3}{2})}_{m}(\cosh r)}. 
\ee
It is easy to check that due to the extra centrifugal term $(D-1)(D-3)/4r^2$,
the corresponding potentials in $D$ dimensions are not shape invariant except
when $D=3$. However, if we consider the approximate extended P\"oschl-Teller
potentials as discussed in Sec. 4.2.1, then as we now show, one gets shape
invariant potentials with translation. In that case
the corresponding superpotential $(W(r)=-\frac{d}{dr}[\ln\chi^{(-)}_{0,m}(r)])$ is given by
\be
W(r)=W_{1}(r)+W_{2}(r)
\ee
Where
\be
W_{1}(r)=-\frac{\phi'_{0}(r)}{\phi_{0}(r)}\quad \mbox{and} \quad  W_{2}(r)=-\frac{\phi'_{m}(r)}{\phi_{m}(r)}.
\ee
Using $W(r)$, the partner potential $V^{(-)}_{eff,m}(r)$ is same as given in 
Eq. (\ref{pot4}) and $V^{(+)}_{eff,m}(r)$ is obtained by using $V^{(+)}_{m}(r)
=W^2(r)+W'(r)$ or simply by replacing $A'\longrightarrow A'-1$ in $V^{(-)}_{eff,m}(r)$. Hence the potentials $V^{(-)}_{eff,m}(r)$ is shape invariant potential 
(with translation) and satisfy
Eq.(\ref{sip}) for any arbitrary values of $D$ and $\ell$. For a check we are giving here some simple cases for $V^{(-)}_{eff,m}(r)$ and $V^{(+)}_{eff,m}(r)$.

{\bf Case (a): m=0 and $\ell\ne0$}

 For $m=0$ and any arbitrary values of $D$ and $\ell$,  Eq. (\ref{pot4}) gives $V^{(-)}_{eff,0}(r)$ as   
\be
V^{(-)}_{eff,0}(r)=(B'^2+A'(A'+1)){\rm cosech^2 r}-B'(2A'+1){\rm cosech r}+A'^2 \coth r
\ee 
and the partner potential $V^{(+)}_{eff,0}(r)$ is given by
\be
V^{(+)}_{eff,0}(r)=(B'^2+A'(A'-1))){\rm cosech^2 r}-B'(2A'-1){\rm cosech r} \coth r+A'^2.
\ee
The potential $V^{(+)}_{eff,0}(r)$ can also be obtained simply by replacing $A'\rightarrow A'-1$ in $V^{(-)}_{eff,0}(r)$ 
and satisfy Eq.(\ref{sip}).
For $D=3$ and $\ell=0$ the parameters $A'\rightarrow A$;\quad $B'\rightarrow B$ and thus these potentials corresponds to the conventional shape invariant 
P\"oschl Teller potentials given in ref. \cite{cks}.
 
{\bf Case (b): $m=1$ and $\ell\ne0$}

For $m=1$ and arbitrary $\ell$, the partner potentials 
\ba
V^{(-)}_{eff,1}(A',B',r)=V^{(A',B')}_{GPT}(r)+\frac{2(2A'+1)}{(2B'\cosh r-2A'-1)}-\frac{2[4B'^2-(2A'+1)^2]}{(2B'\cosh r-2A'-1)^2}+A'^2
\ea
and
\be
V^{(+)}_{eff,1}(A',B',,r)=V^{(A'-1,B')}_{GPT}(r)+\frac{2(2A'-1)}{(2B'\cosh r-2A'+1)}-\frac{2[4B'^2-(2A'-1)^2]}{(2B'\cosh r-2A'+1)^2}+A'^2.
\ee
These potentials satisfy the shape invariant property (\ref{sip}) i.e.,
\be
V^{(+)}_{eff,1}(A',B',r)=V^{(-)}_{eff,1}(A'-1,B',r)+2A'-1.
\ee
 For $D=3$ and $\ell=0$, the above expressions matches exactly with the results obtained in \cite{bqr,rabxm,rab}. 

 \section{Results and discussions}
 
 In the present manuscript by using PCT approach we have generated exactly solvable rationally extended $D$-dimensional 
 radial oscillator and GPT potential and constructed their bound state wavefunctions in terms
of $X_m$ exceptional Laguerre and Jacobi orthogonal polynomials respectively. The extended potentials are isospectral to their conventional
counterparts. For the oscillator case we have shown that the rationally 
extended $D$-dimensional oscillator potentials are shape invariant with 
translation. For the Jacobi case, this is not true unless $D=3$. For $l \ne 0$
 we also obtained approximate extended GPT potentials and these have been 
shown to be shape invariant. 
For the particular case ($D=3$) the potentials corresponds to the potentials 
obtained by Quesne et.al \cite{que,bqr} and others and thus provide a powerful 
check on our calculations.

  {\bf Acknowledgment}

One of us (NK) acknowledges financial support from UGC under the BHU-CRET fellowship.

\end{document}